\documentclass[twocolumn]{aastex62}

\newcommand\msun{\rm M_\odot}
\graphicspath{{./}}

\received{28 July 2022}
\accepted{15 August 2022}
\submitjournal{ApJ}

\shorttitle{OBe Stars as Post-SN Runaways}
\shortauthors{Dallas, Oey \& Castro}

\begin{document}
\title{CLASSICAL OBe STARS AS POST-SUPERNOVA RUNAWAYS: CONFIRMING BINARY ORIGINS}

\correspondingauthor{M. S. Oey}
\email{msoey@umich.edu}

\author[0000-0001-9692-9751]{Matthew M. Dallas}
\affil{Department of Astronomy, University of Michigan, 
1085 South University Ave., Ann Arbor, MI 48109-1107, USA \\
}
\affil{Present address:  Space Telescope Science Institute, 3700 San Martin Drive, Baltimore, MD   21218}

\author[0000-0002-5808-1320]{M. S. Oey}
\affil{Department of Astronomy, University of Michigan, 
1085 South University Ave., Ann Arbor, MI 48109-1107, USA \\
}

\author[0000-0003-0521-473X]{Norberto Castro}
\affiliation{Leibniz-Institut für Astrophysik Potsdam (AIP), An der Sternwarte 16, 14482, Potsdam, Germany}

\begin{abstract}
Massive binaries play an important role in fields ranging from gravitational wave astronomy to stellar evolution. We provide several lines of evidence that classical OBe stars in the Small Magellanic Cloud (SMC) obtain their rapid rotation from mass and angular momentum transfer in massive binaries, which predicts that the subsequent supernovae should often eject OBe stars into the field.  We find that (1) OBe stars have a higher field frequency than OB stars; (2) our cumulative distribution function (CDF) of stellar distances from O stars shows that OBe stars are indeed much more isolated than ordinary OB stars of corresponding spectral types; (3) the CDFs of OBe stars approach that of high-mass X-ray binaries (HMXBs), which are confirmed post-supernova objects; and (4) Oe stars are as isolated from clusters as Be stars, implying that their final masses are relatively independent of their initial masses, consistent with major mass transfer. Lastly, we also find that the spatial distribution of supergiant OBe stars differs from that of classical OBe stars, consistent with the different mechanism responsible for their emission-line spectra.
\end{abstract}

\keywords{Be stars --- Oe stars --- core-collapse supernovae --- stellar evolutionary models --- high-mass X-ray binary stars --- circumstellar disks --- interacting binary stars --- multiple star evolution --- compact objects --- star clusters --- Small Magellanic Cloud --- runaway stars}

\section{Introduction} \label{sec:intro}

Classical OBe stars are non-supergiant OB stars that exhibit Balmer emission lines in their spectra, first observed in 1866 \citep{Secchi1866}.
The emission lines result from circumstellar disks that are expelled by 
near-critical stellar rotation \citep[e.g.,][]{Rivinius2013}, and 
how the stars obtained their fast rotation has not been not well understood.  
The transfer of mass and angular momentum in binary interactions \citep[e.g.,][]{Kriz, Pols}
is a model that has recently gained much traction. The vast majority of OB stars are binaries \citep[e.g.,][]{Sana, Chini}, and binary population synthesis models produce fast-rotating populations consistent with the parameters of the observed Be star population in the Milky Way \citep{Shao, Boubert}. Moreover, this model predicts that Be stars should often have post main-sequence binary companions, and these have been observationally verified for a number of objects \citep[e.g.,][]{Wang, Klement2019, Bodensteiner}. 

In the binary model for OBe stars, the more massive primary fills its Roche lobe and becomes a mass donor to the companion, thereby increasing the mass gainer's angular momentum enough to generate the decretion disk \citep[]{Pols}. Massive donors later explode as supernovae (SNe), accelerating the mass gainers and often unbinding them from star clusters \citep[e.g.,][]{Blaauw1961}. This type of ejection is dubbed the binary supernova scenario \citep[BSS; ][]{Hoogerwerf}.

Be stars have higher proper motions than B stars \citep[]{Berger}, supporting this scenario. Furthermore, the number of OBe field stars in the Small Magellanic Cloud (SMC) is consistent with the number of expected BSS ejections \citep[]{Johnny}, and almost all  high-mass X-ray binaries (HMXBs) in that galaxy are emission line stars \citep[e.g.,][]{Maravelias}. Recently, \citet{Hastings21} find that the frequency of OBe stars in clusters is consistent with binary mass-transfer origins. However, the possibility also remains that some OBe stars originate from a different mechanism \citep[e.g.,][]{Langer98}. 

Here, we further examine whether massive OBe stars are largely post-SN objects. If so, their enhanced transverse velocities \citep[e.g.,][]{Renzo} would cause them to be
more isolated than they would be in single-star models for the OBe phenomenon \cite[e.g.,][]{Ekstrom2008}. \citet[][hereafter ST15]{ST} used this principle to demonstrate that luminous blue variables are likely mass gainers that are "kicked" into the field by BSS ejections. 
We apply the same spatial analysis used by ST15 to test whether OBe stars systematically avoid clusters compared to non-emission line stars. Following ST15, we compile the cumulative distribution function (CDF) of the projected separations between OBe stars and their nearest O stars, and we compare with the non-OBe stars. Since we expect O stars to not travel far from their birth clusters, the distance to the nearest O star effectively measures the relative isolation of the target star.

\section{The SMC OB Star Sample} \label{sec:Sample Description}

We use the \citet[][hereafter OKP]{OKP} sample of 1360 OB stars in the SMC with spectral types earlier than $\sim$B2. This sample is photometrically selected, based on the $UBVI$ survey of \citet[]{Massey2002}, and it is spatially complete over most of the star-forming body of the SMC. The field component is largely complete for masses $\gtrsim 20\ \rm M_\odot$ \citep{Lamb2013, RIOTS4}, with a higher limit for clusters. Existing OB spectral types are obtained from spectroscopic data compiled in the SIMBAD database, primarily from the SMC surveys of \citet[][RIOTS4]{RIOTS4}, \citet{Massey2002}, and \citet{Evans2004}. In cases where there were two conflicting spectral types reported for a given star, we generally adopted the RIOTS4 value if available, and otherwise the type with the highest spectral resolution was chosen; 
if more than two spectral types were found, the most frequently identified type was adopted.
Stars with no published spectral type are retained as OB candidates.The adopted spectral types are given in Table~\ref{tab:Catalog}.

\begin{deluxetable*}{ccclccccll}
\tablecaption{OKP Sample} 
\tablewidth{0pt}
\tablehead{
\colhead{[M2002]\tablenotemark{a}} & \colhead{RAdeg (J2000)} & \colhead{DEdeg} & \colhead{SpType\tablenotemark{b}} & \colhead{Field/Group\tablenotemark{c}} & \colhead{OBe\tablenotemark{d}} & \colhead{$m_{\rm up}\ (M_\odot)$\tablenotemark{e}} & \colhead{$m_{\rm low}\ (M_\odot)$\tablenotemark{e}}
}

\startdata
  107 & 10.1171 & --73.5425 & Be3               &    F & e      & 48 & 48\\  
   298 & 10.1832 & --73.4063 & B1e3+             &   F & e     & 17 & 17\\
  1037 & 10.3880 & --73.4256 & B0.5 V             &   F &   \nodata   & 14 & 14\\   
  1600 & 10.5417 & --73.2323 & O8.5 V / Be?       &   F & e?    & 22 & 22\\  
  1631 & 10.5515 & --73.3866 & B1e2                &   F &  e   & 22 & 22\\   
\enddata
\tablenotetext{a}{ID from \citet{Massey2002} SMC catalog.}
\tablenotetext{b}{Spectral types obtained as described in the text.RIOTS4 OBe classifications include Arabic numerals corresponding to \citet{Lesh1968} emission classes following the "e" designation.
}
\tablenotetext{c}{Field or Group membership (see text), denoted as F and G, respectively.}
\tablenotetext{d}{OBe and candidate OBe stars are indicated with "e" and "e?", respectively. Note that OBe supergiants are also identified.} 
\tablenotetext{e}{Empty values are those where our method was unable to obtain a mass estimate.}
\tablecomments{Table~\ref{tab:Catalog} is published in its entirety in machine-readable format. A portion is shown here for guidance regarding its form and content.}
\label{tab:Catalog}
\end{deluxetable*}

OKP OBe stars not identified from the main spectroscopic surveys above were mostly identified from \citet[]{Meyssonnier1993}, who carried out a spatially complete, objective prism survey of the entire SMC, targeting H$\alpha$ emission-line objects down to a photographic magnitude of 18 in the continuum.
We also utilize \citet[]{Mennickent2002}, who identified Be candidates primarily from photometric variability. If a star was listed in these catalogs
as an emission-line object, and was given a spectral type of O or B in a different catalog, it was included in our certain or candidate Oe and Be star lists and identified as "Em*" \citep{Meyssonnier1993} and/or "Be?" \citep{Mennickent2002} in Table~\ref{tab:Catalog}.

\citet{Aadland2018} discuss how the nearest O-star CDFs for different OB spectral classes can depend on the completeness of spectral type availability. Since \citet[]{RIOTS4} obtained spectra of nearly all the field stars, the parent OKP sample has a strong bias for available spectral types in the field relative to groups. 
Appendix~\ref{sec:completeness} reviews the completeness of the OKP sample relative to other SMC OBe and OB surveys.
However, we stress that what matters here is whether there is a {\it relative} bias against OKP OBe detections in clusters compared to field. Comparison with the Evans surveys show no evidence of such a bias, especially considering that \citet{Evans2006} specifically considers the rich clusters NGC 330 and 346 (Table~\ref{tab:Completeness}). 
Indeed, OBe stars are more easily detected than OB stars since they are unlikely to be newborn, embedded objects, and therefore have enhanced luminosities relative to the ZAMS.  Moreover, the disk emission contributes to their optical fluxes and provides easily-detected line emission.  Thus we caution that these effects generate a bias favoring the detection of OBe stars, potentially setting a lower mass selection than for OB stars.  On the other hand, general incompleteness in a few highly reddened regions may dilute this selection bias, but this mainly applies to the densest clusters like NGC~330 \citep[e.g.,][]{BodensteinerN330} and NGC~346.

\begin{deluxetable*}{llcccccccc}
\tablecaption{Populations of SMC OB stars}
\tablewidth{0pt}
\tablehead{
 & & \multicolumn{4}{c}{Numbers in Populations\tablenotemark{a}} & \multicolumn{3}{c}{Nearest O-Star Separations (pc)\tablenotemark{b}} \\
\colhead{Row} & \colhead{Spectral Type} & \colhead{Field} & \colhead{Groups} & \colhead{Total} & \colhead{Proportion in Field}
& \colhead{Median} & \colhead{ Mean} & \colhead{Std Err}
}
\startdata
(1) & Early O (O3--O7) & 15 & 39 & 54 & $0.28\pm0.08$ & 19 & 28 & 5.1 \\ 
(2) & Late O (O8--O9) & 60 & 75 & 135 & $0.44\pm0.07$ & 28 & 40 & 4.2 \\
(3) & B (B0--B2) & 91 & 102 & 193 & $0.47\pm0.06$ & 30 & 45 & 4.4 \\
(4) & OB\tablenotemark{c} & 11 & 383 & 394 & $0.03\pm0.01$ & 13 & 21 & 1.6  \\
(5) & Oe & 22 & 16 & 38 & $0.58\pm0.16$ & 34 & 56 & 9.4 \\
(6) & Be & 115 & 108 & 223 & $0.52\pm0.06$ & 39 & 63 & 4.9 \\
(7) & OBe\tablenotemark{c} & 11 & 107 & 118 & $0.09\pm0.03$ & 21 & 30 & 2.6 \\ 
(8) & OB or OBe\tablenotemark{c} & 5 & 24 & 29 & $0.17\pm0.08$ & \nodata & \nodata & \nodata \\
(9) & O,B HMXB\tablenotemark{d} & 4 & 3 & 7 & $0.57\pm0.36$ & 54 & 54 & 5.5 \\
(10) & OBe HMXB & 19 & 20 & 39 & $0.49\pm0.14$ & 47 & 61 & 8.1 \\
\hline %
(11) & {Total confirmed O\tablenotemark{e,f} }& 81 & 121 & 202 & $0.40\pm0.05$ & 26 & 37 & 3.1 \\
(12) & Total OB\tablenotemark{f,g} & 174 & 599 & 773 & $0.23\pm0.02$ & 22 & 36 & 1.8\\
(13) & Total OBe\tablenotemark{g} & 167 & 251 & 418 & $0.40\pm0.04$ & 36 & 57 & 3.3\\
(14) & Total OB and OBe\tablenotemark{f,h} & 346 & 874 & 1220 & $0.28\pm0.02$ & \nodata & \nodata & \nodata   \\ 
(15) & {Total HMXB\tablenotemark{i}} & 24 & 25 & 49 & $0.49\pm0.12$ & 48 & 58 & 6.5\\ 
\hline
(16) & Cross-class binaries\tablenotemark{f} & 7 & 3 & 10 & $0.47\pm0.21$ & \nodata & \nodata & \nodata  \\
(17) & O,B I/II & 21 & 59 & 80 & $0.26\pm0.06$ & 24 & 42 & 4.9 \\
(18) & OBe I/II\tablenotemark{d} & 6 & 35 & 41 & $0.15\pm0.06$ & 29 & 46 & 7.7 \\
\enddata 
\tablenotetext{a}{OKP stars with spectral type $\leq$ B2.  Supergiants (luminosity class I/II) are excluded from all categories except in rows 15, 17, and 18.  HMXBs are excluded from rows 1--8, but included in rows 11--15.  "Group" stars are non-field stars.}
\tablenotetext{b}{Listed 
values are calculated for the CDFs shown in Figure~\ref{fig:f2}a.}
\tablenotetext{c}{"OB", and "OBe" stars have uncertain classification between Early vs Late O, O vs B, and Oe vs Be, respectively. "OB or OBe" have uncertain emission-line status.  There are no published spectral types for 373 "OB" stars and 107 "OBe" stars. }
\tablenotetext{d}{Includes one star of uncertain emission-line status.}
\tablenotetext{e}{Includes 5 Field and 7 Group stars of uncertain Early vs Late O status from row 4, and 1 Field O HMXB from row 9.}
\tablenotetext{f}{For our analysis, binaries with two stars of the same classification are treated as a single star of the relevant class, as are binaries with a member not meeting our selection criteria.
Ten binaries in row 16
have components individually included in rows 1--4, but treated as single stars in rows 11, 12 and 14.}
\tablenotetext{g}{Total number excluding row 8.}
\tablenotetext{h}{Total number including row 8.} 
\tablenotetext{i}{All OKP HMXBs, including 1 B[e] HMXB and 2 supergiant HMXBs, which are excluded from rows 9--14.  These all represent post-SN objects.}
\label{tab:Populations}
\end{deluxetable*}

We sorted the stars into the categories
shown in Table~\ref{tab:Populations}, separating out supergiants (luminosity classes I and II).
We exclude 16 other stars such as B[e] stars, WR stars, and stars with spectral types $>$ B2. Our final sample size, omitting these 16, is 1344 stars. 
Tables~\ref{tab:Catalog} and \ref{tab:Populations} indicate field and group stars, with groups defined as having at least 3 stars associated by the friends-of-friends algorithm for a clustering length of $l_c=28$ pc \citep{OKP}.  Single stars and those with only one other OB star within $l_c$ are defined here to be field stars.  We adopt an SMC distance modulus of 18.9 \citep{Harries2003}.

\begin{figure*}
\plotone{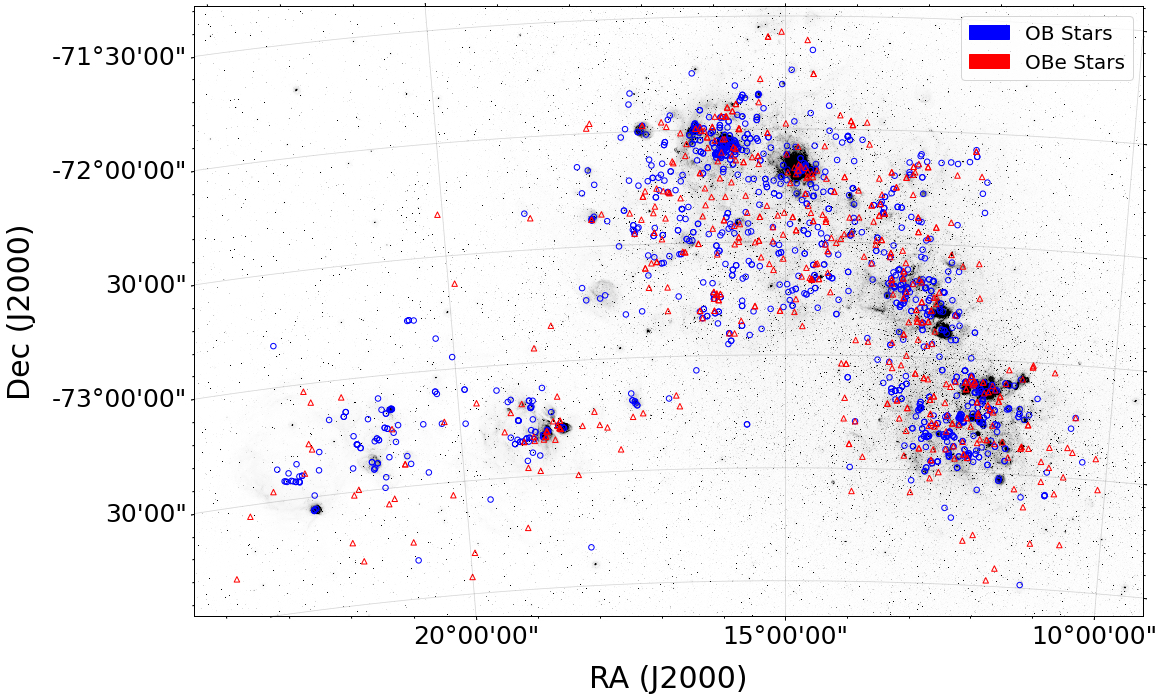}
\caption{OB (blue) and OBe (red) stars from OKP plotted on an H$\alpha$ image of the SMC \citep[]{Halphamap}. The 29 stars of uncertain emission-line status
in Table \ref{tab:Populations} are included here as OBe stars. }
\label{fig:f1}
\end{figure*}

We see that of the total  418 OBe stars, $40\%\pm4$\% occur in the field, despite the field star sample only making up $28\%\pm2$\% of the sample. 
Figure \ref{fig:f1} also shows that OBe stars have a greater spatial extent than OB stars, extending to larger distances from the star-forming body of the SMC.

\section{Spatial isolation of OBe stars} \label{sec:Results}

\begin{figure*}
\plotone{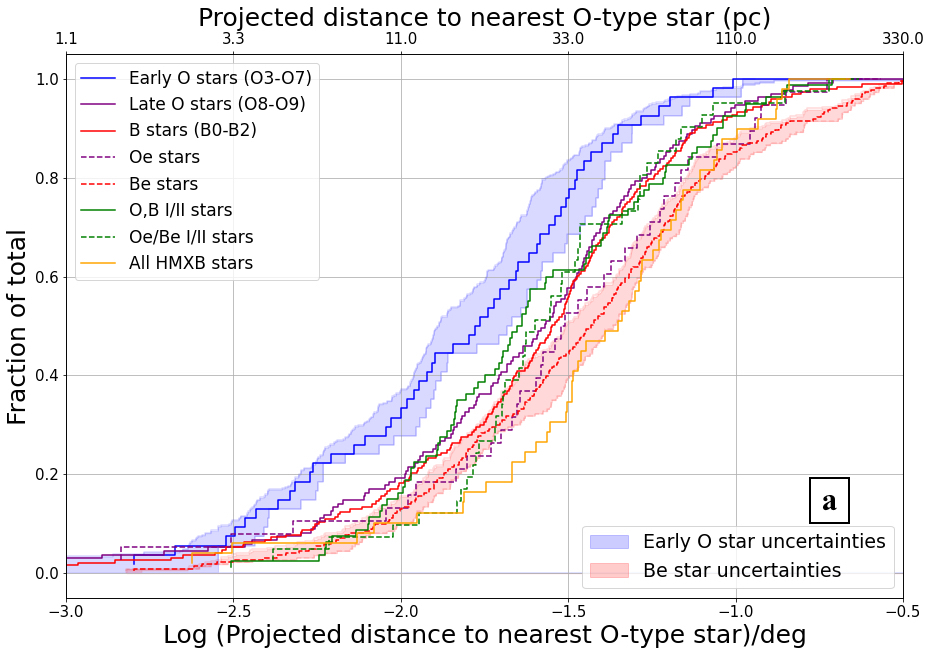}
\plottwo{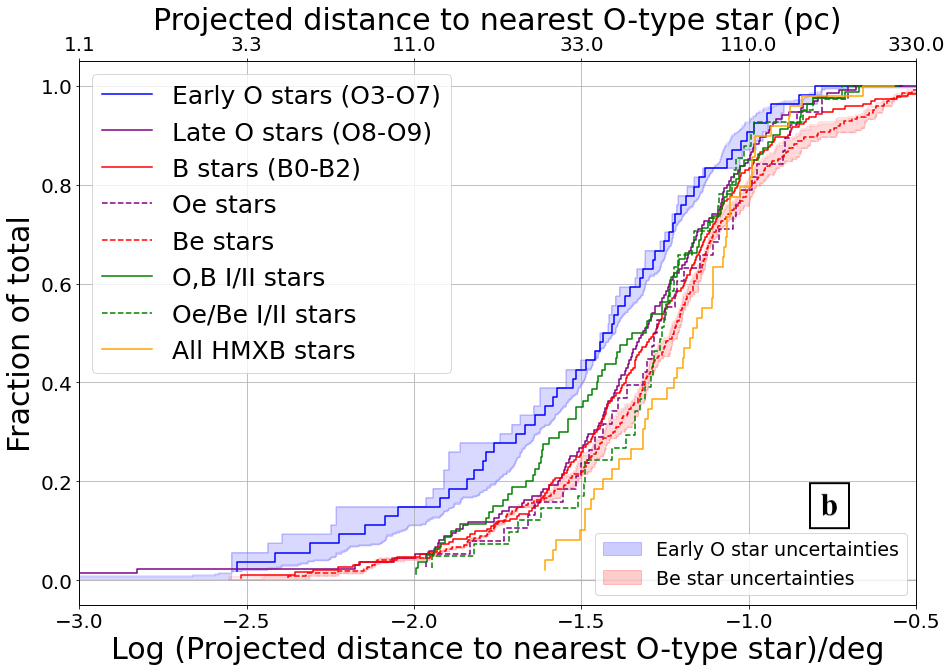}{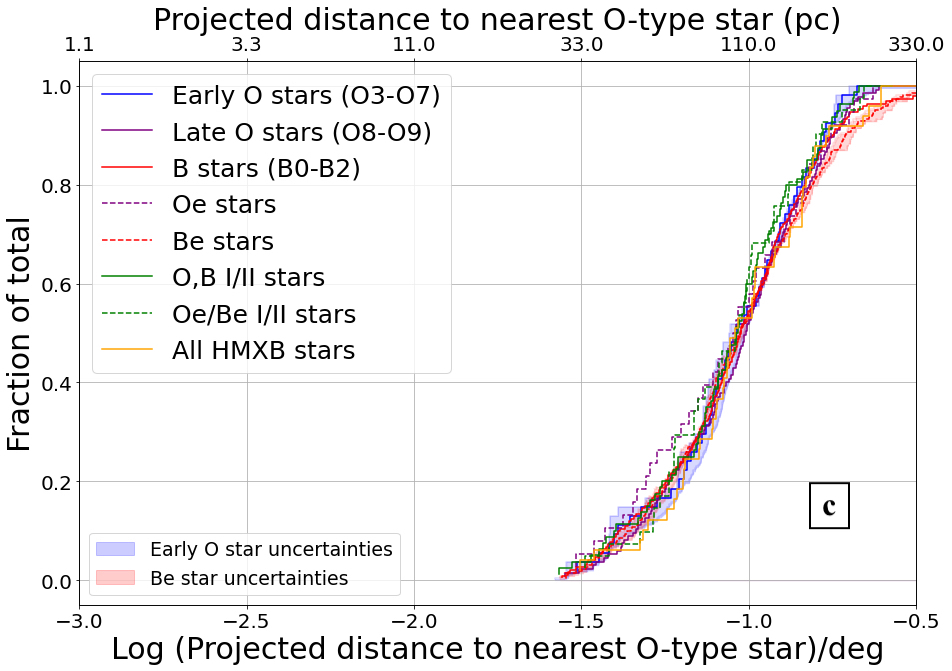}
\caption{Panel (a):  Cumulative distribution functions of distances to the nearest O-type star for the shown populations. 
The top and bottom axes show separation in pc and degrees, respectively. The maximum and minimum CDFs from which the median is constructed is shown by the shaded regions for Early O stars and Be stars (see text).
Panel (b):  Only allowing confirmed O stars as candidate nearest O stars.  Panel (c): Only allowing field O stars, including uncertain OB stars, as candidate nearest O stars.}
\label{fig:f2}
\end{figure*}

Figure \ref{fig:f2}a plots the CDFs of separations from the nearest O-stars that are not Oe, HMXB or supergiant, for different stellar classes in Table~\ref{tab:Populations}.  
These are obtained by first constructing the
"maximum" CDF using all stars in a given population, including those with uncertain spectral types; and second, the "minimum" CDF using only stars with spectral types that are not reported to be uncertain. This range is shown by the shaded regions for the O and Be star populations as representative examples. 
The adopted CDFs are then the medians between these extremes.
Binary stars with members in two different populations are included in both when calculating the CDF. 
Table~\ref{tab:accounting} gives the numbers and detailed accounting of the membership in the OB and OBe CDFs,and Table~\ref{tab:Populations} gives the separations from nearest O stars for each category as obtained from the CDFs.

Table \ref{tab:Populations} shows that almost all the OB stars with uncertain classifications are in groups, and moreover, they constitute almost half of our entire OB sample.  
However, we note that the OKP sample is based on uniform photometric selection of OB stars, and the uncertainty in the spectral classifications  is accounted for by our method based on the maximum and minimum CDFs. Our sample does include 63 stars with spectral types identified by ranges such as "B0-5" or "B1-3", meaning that they that may be later than the nominal limit of B2. Excluding these from our sample results in an insignificant offset to the CDF medians. For B, Be, and OB supergiants, their removal affects the median by less than 1.0 pc; and for HMXBs and OBe supergiants, the median changes by 2.4 pc and 1.4 pc, respectively.  Thus, the inclusion of these stars does not make an appreciable difference to our results. 

The CDFs in Figure~\ref{fig:f2}a allow all of the OB stars with uncertain classifications (Table~\ref{tab:Populations}, line~4) to serve as "home base" O stars, i.e., candidate nearest O stars.  
Note that these do not necessarily represent the parent clusters of the target stars, since the clustering length for OB stars is only 28 pc \citep{OKP}; thus the CDF should simply be regarded as a measure of relative isolation.
If we allow only confirmed O stars (Table~\ref{tab:Populations}, line~11) to serve as "home base" stars, we obtain the CDFs in Figure~\ref{fig:f2}b.  
This can also be compared with Figure~\ref{fig:f2}c, which allows only field O stars, including "OB" stars, to serve as "home base" stars, thereby forcing all populations to follow the field star CDF.
Since almost all of the uncertain "OB" stars belong to groups, we see that Figure~\ref{fig:f2}b effectively removes many groups, which moves the CDFs for some populations farther into the field. Therefore, Figure~\ref{fig:f2}a is more realistic than Figure~\ref{fig:f2}b.

Figure \ref{fig:f2}a indeed shows the expected progression of 
early O stars being the least isolated, followed by 
late O stars, and then B stars, confirming the trend reported by ST15. However, we also immediately see that both Oe and Be stars are farther out in the field than their non-emission-line counterparts. Moreover, Oe and Be stars are farther in the field than even evolved supergiant OB stars. 

These trends are supported by Mann-Whitney statistics, which test for difference in location of distributions.  The $p$-values for comparing the Late O vs Oe and B vs Be CDFs are 0.089 and 0.005, respectively.  In contrast, $p > 0.5$ for both Late O vs B, and Oe vs Be.
The trends are corroborated by our findings above that OBe stars have both a higher field star frequency and greater spatial distribution than their OB counterparts (Section~\ref{sec:Sample Description}).

The Oe and Be CDFs in Figure~\ref{fig:f2}a are statistically indistinguishable ($p = 0.83$), in contrast to the corresponding OB distributions.
This implies that on average, Be stars do not drift further into the field than Oe stars, contrary to expectations based on the relative lifespans of O and B stars. 
This is consistent with the mass-transfer scenario for OBe stars, whereby the masses of the gainers are altered by varying amounts of mass transfer, which can also affect both the spectral type and nuclear burning lifetime, diluting the expected relationship between spectral type and age, and therefore, field distance.

Thus, these results support the scenario that OBe stars originate from BSS ejections in massive binary systems.
This is consistent with the findings of \citet{Johnny}, who find that the field OBe population statistics and kinematics are broadly consistent with BSS origins. The effect may also be amplified by the two-step ejection process \citep{Pflamm2010} discussed in that work.

\begin{figure*}
\plottwo{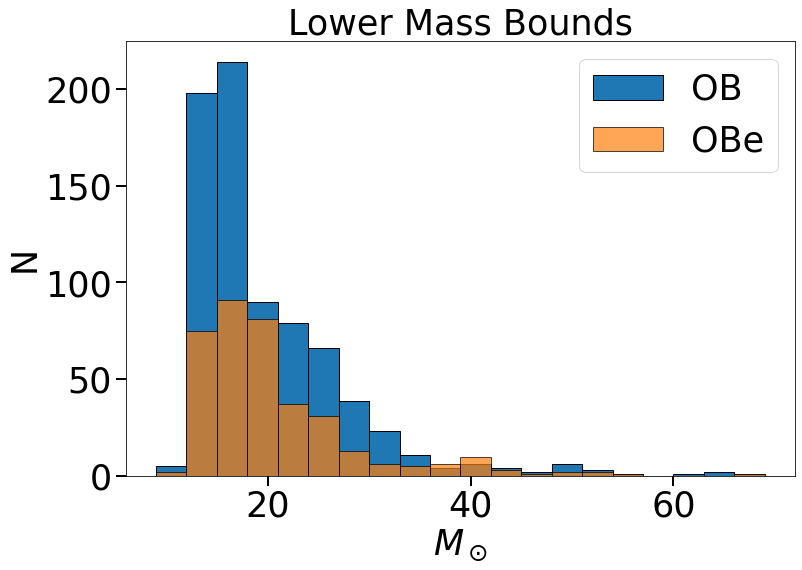}{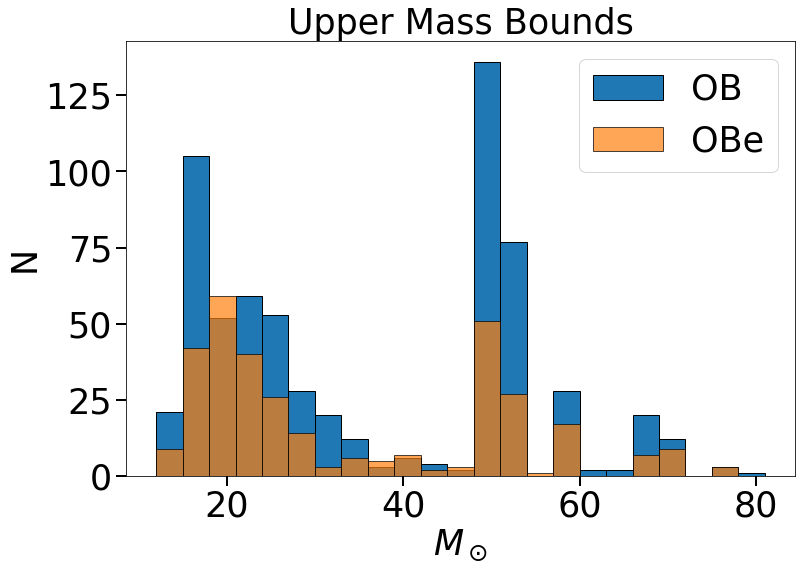}
\caption{Mass distributions when including the stellar lower mass limits (left) versus upper mass limits (right) for stars with uncertain classifications spanning a range of spectral type.  The bimodality for the upper mass estimates results from the fixed adopted spectral-type ranges (see text) which translate into a limited range of upper-mass limits, most of which overestimate the true masses.
These mass estimates show that our sample of OBe stars are not significantly biased toward lower masses.}
\label{fig:MassDistrib}
\end{figure*}

\begin{figure}
\plotone{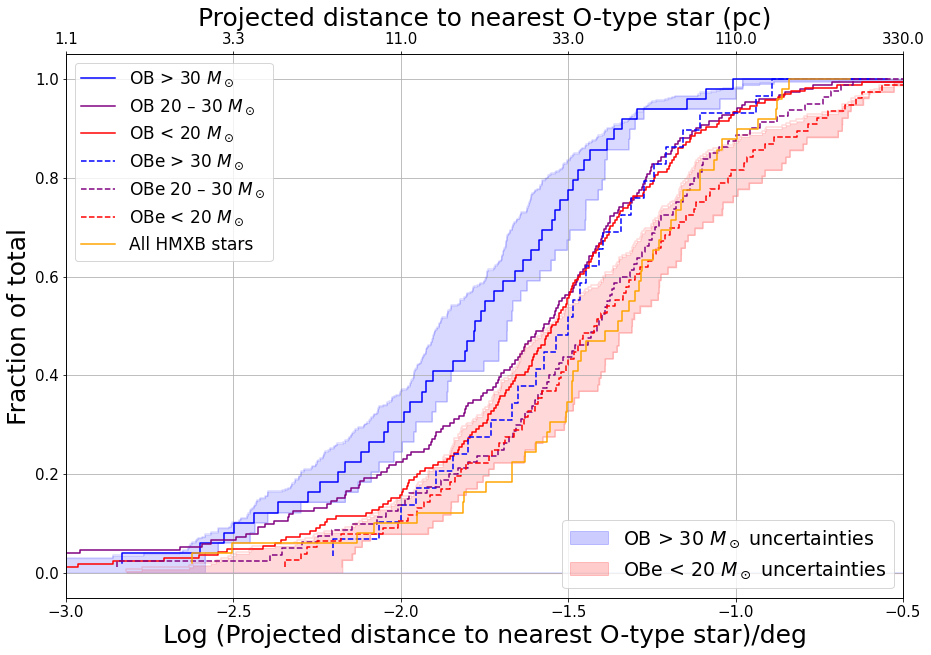}
\caption{
Cumulative distribution functions of distances to the nearest O-type star for the shown populations similar to Figure~\ref{fig:f2}, but with OB and OBe populations binned by stellar mass as shown. Representative maximum and minimum CDFs from which the medians are constructed are shown by the shaded regions for OB stars with $m > 30\ \msun$ and OBe stars with $m < 20 \msun$.  }
\label{fig:CDFmass}
\end{figure}

We caution that certain effects may complicate our findings. 
First, the CDF trend could be mimicked if the OBe stars originate from systematically lower ZAMS masses than the OB stars. Since OBe stars in the SMC tend to be slightly older than other OB stars \citep{Martayan2007b} and have decretion disks, they appear slightly redder, which may introduce a slight bias toward selecting lower-mass objects with our photometric selection criteria.
Second, there are likely many binary mass gainers and quiescent OBe stars without currently active decretion disks, which appear in the OB CDF.  The two effects counteract each other, so it is unclear how they affect our final results.

We therefore also construct CDFs similar to Figure~\ref{fig:f2}, but for OB and OBe stars binned by mass ranges instead of spectral type.  The masses are obtained using photometry from \citet{Massey2002} and the spectral types compiled above.  We convert from spectral type to $T_{\rm eff}$ following \citet[][see also Lamb et al. 2013]{Castro2021}; luminosities are obtained from $\log g$ based on the photometry, $T_{\rm eff}$, and extinction curve of \citep{Fitzpatrick2007}.  
The masses (Table~\ref{tab:Catalog}) are obtained by interpolating the positions of the stars on the H-R diagram using the \citet{Brott2011} evolutionary tracks for SMC metallicity.  Stars with ambiguous spectral types of simply "O", "B", and "OB" are assigned ranges of O3 -- O9, B0 -- B3, and O3 -- B3, respectively.  This causes an artificial bimodality in the resulting mass distribution (Figure~\ref{fig:MassDistrib}) because most of these ranges allow spectral types as early as O3 for any given star.  In reality, the vast majority of stars with these uncertain types must have spectral types near the late end of their ranges, as dictated by the initial mass function.  Thus, the number of high-mass stars is significantly overestimated.
The main point of the comparison however, is to demonstrate that there is no signficant offset in the OBe masses relative to those OB stars, which indeed appears to be the case.

For completeness, Figure~\ref{fig:CDFmass} shows the resulting CDFs for $m> 30\ \msun$, $20 < m \leq 30\ \msun$, and $m \leq 30\ \msun$.  As in Figure~\ref{fig:f2}a, the adopted CDFs are the medians between the maximum CDF, which uses all stars in the given mass range, including those which have only an upper or lower limit within the range, or where a mass estimate was unable to be obtained; and the minimum CDF, which uses only stars for which both the upper and lower mass limits are within the given range. These CDFs are therefore constructed analogously to those in Figure~\ref{fig:f2}a.  Representative examples of maximum and minimum CDFs are shown in the shaded regions for the OB stars with $> 30\ \msun$ and OBe stars having  $< 20\ \msun$.  As a result of the overestimated upper masses for stars with uncertain spectral types, the maximum CDFs tend to be overpopulated and therefore the median CDFs are biased toward smaller field distances, especially for the highest-mass population.  A similar, but less pronounced, effect therefore applies to Figure~\ref{fig:f2}.
Overall, Figure~\ref{fig:CDFmass} confirms the trends in Figure~\ref{fig:f2}, including the trend that OBe stars are at larger field distances than OB stars in the same mass ranges.

\subsection{Comparison with HMXBs}

If OBe stars are BSS-ejected objects, then their CDFs should reside at a locus similar to that of other known BSS populations. We therefore consider the HMXBs in our sample \citep[]{Haberl}, which are known post-SN objects. 
Table~\ref{tab:Populations} shows that 39 of the 49 HMXBs are confirmed OBe stars, thus the comparison is not independent, but this demonstrates a close link between OBe stars and HMXBs.

Figures \ref{fig:f2}a, \ref{fig:CDFmass}  and Table~\ref{tab:Populations} compare the CDFs for the non-HMXB Oe and Be stars, and all the HMXBs, since there are so few non-OBe HMXBs.
The similarity of the Oe and Be CDFs with HMXBs ($p=0.28$ in Figure~\ref{fig:f2})
supports the scenario that post-SN systems may dominate the classical OBe star population \citep{Vinciguerra2020}. This is consistent with \citet[]{Bodensteiner}, who find no main-sequence companions for any Be systems in the Milky Way. 

Figures~\ref{fig:f2}a, \ref{fig:CDFmass} and Table~\ref{tab:Populations} are also consistent with HMXBs being the population furthest into the field,
consistent with expectations that they represent the tightest binaries, capable of surviving the supernova kicks.  Models show that they are more strongly accelerated than unbound objects  \citep[][]{Renzo, Podsiadlowski}.

Comparing Figures~\ref{fig:f2}a and \ref{fig:f2}b, we see that HMXBs are especially sensitive to the removal of groups, implying that they originate from sparser OB groups than the other populations.  This suggests that they originate from more evolved groups, and therefore may correspond to lower-mass, rejuvenated progenitors. This needs to be confirmed, but is consistent with expectations that HMXBs
dominate at lower progenitor masses at lower, SMC metallicity
than at high metallicity \citep[e.g.,][]{Heger2003}. 

\section{Discussion} \label{sec:Discussion}

The relative isolation of OBe stars supports the model whereby these emission-line stars form through massive binary interaction, since
the companion gains both mass and angular momentum \citep[e.g.,][]{Pols}. If the mass gainer approaches its critical rotation velocity as a consequence of stellar evolution \citep[e.g.,][]{Zhao2020}, a decretion disk forms, and the gainer becomes an OBe star. The donor continues to evolve, likely becoming a stripped, He-burning star, and subsequently a core-collapse SN, kicking the system.
The final end state of non-disrupted massive binaries may often be double compact systems. 
The high frequency of massive OBe stars in the SMC implies
that mass transfer in our objects is dominated by Case A or B, and unlikely to have significant contributions at later evolutionary stages. This is also consistent with Be X-ray binary population synthesis models \citep{Vinciguerra2020}.  

The scenario that most OBe stars represent objects that have been spun up by mass transfer predicts that we should expect to see pre-SN examples of this interaction.
Two of our OBe stars, [M2002]SMC-30744 and 41095, are double-lined spectroscopic binaries, one of which is also an eclipsing binary.  
These interacting binaries are not classical OBe stars, and thus excluded from the OBe sample.  But
they may represent examples of actively accreting systems that will evolve into classical OBe stars.

We note that OBe supergiants are found at the same distances as OB supergiants (Mann-Whitney $p=0.33$), and not with the other OBe stars (Figure~\ref{fig:f2}a, Table~\ref{tab:Populations}).
This is consistent with an evolved population that has a different origin for the emission lines, which are believed to form in the stellar winds rather than a decretion disk \citep{Puls2008}.

\section{Conclusion} \label{sec:Conclusion}

Binary mass transfer is increasingly seen as the mechanism for 
spinning up classical OBe stars, enabling formation of their decretion disks.
In particular, previously hidden subdwarf companions are now observed for lower-mass Be stars \citep[e.g.,][]{Wang}, binary population synthesis models produce Be star populations that are consistent with observations \citep{Shao, Boubert}, and a lack of main-sequence companions to massive Be stars appears to be established \citep[]{Bodensteiner}. Moreover, the numbers of OBe stars seem generally consistent with post-SN binary origins \citep{Johnny}.

Using the spatially complete sample of 1344 OB stars in the SMC from \citet{OKP}, we show that the spatial distribution of OBe stars further confirms that these objects experienced BSS ejections, as follows.
(1) We find that our field stars correspond to $40\%\pm4$\% vs $28\%\pm2$\% of OBe and OB stars, respectively, consistent with the visual impression of their relative spatial distribution (Figure~\ref{fig:f1}).
Using the CDF of stellar distances to nearest O stars to evaluate relative isolation for different populations, we find that, (2) 
OBe populations are more isolated than their counterpart OB populations (Figure~\ref{fig:f2}a, Table~\ref{tab:Populations}).
Moreover, (3) OBe stars reside at distances into the field similar to that of HMXBs, which are known post-SN binaries. Finally, (4) the Oe and Be-star CDFs occupy the same locus, contrary to their spectroscopic life expectancies. This implies that their observed spectral types are inconsistent with their birth masses, again supporting the mass-transfer scenario. Thus, \textit{several lines of evidence} indicate that OBe stars largely originate as mass gainers in close, post-SN, massive binary systems.

We also find that OBe supergiants are not as isolated as classical OBe stars, as expected from their different origin.

\acknowledgments
We thank Grant Phillips, Irene Vargas-Salazar, Johnny Dorigo Jones, and Lena Komarova for help and discussions. We also thank Max Moe, Mathieu Renzo,
and the anonymous referees 
for discussions and comments that greatly improved this paper.  This work was supported by the National Science Foundation, grant AST-1514838 to M.S.O., who also acknowledges office space hospitality from MDRS, LLC. N.C. acknowledges funding from the Deutsche Forschungsgemeinschaft (DFG), CA 2551/1-1.  The Python package astropy was integral to the coding in this research \citep[]{Astropy}.  

\clearpage

\appendix 
\section{Sample completeness}\label{sec:completeness}

Table~\ref{tab:Completeness} examines the completeness of the OKP sample relative to other SMC OBe and OB surveys,
applying OKP selection criteria of spectral type $\leq$ B2, luminosity class III--V, and $B\leq15.21$ or $V\leq15.5$.  We see that we recover $>50$\% of Oe stars in the \citet{Martayan2007, Martayan2010} OBe surveys, which target 85 SMC clusters.  Overall, \citet{Schootemeijer2021} estimate that the SMC has a total of $\sim600$ O-stars.  Including Oe stars, OKP has 239 confirmed O-stars and 778 confirmed plus candidate O-stars
(Table~\ref{tab:Completeness}); thus our O-star sample is also largely complete.
For comparison, the \citet{Bonanos2010} spectroscopic catalog has 250 O-stars applying the OKP criteria {\it without} the magnitude cut.

\section{Subsample membership}

Table~\ref{tab:accounting} gives a detailed accounting of the membership for the "maximum" CDFs constructed for Figure~\ref{fig:f2}, i.e., each spectral-type subsample that includes stars with uncertain spectral types that include possible membership in the given population.  The corresponding "minimum" CDF memberships can be found in Table~\ref{tab:Populations}.

\begin{deluxetable*}{lcccccc}
\tablecaption{ Catalog Comparisons\tablenotemark{a}} 
\tablewidth{0pt}
\tablehead{
\colhead{Populations} & \colhead{OKP\tablenotemark{b}} & \colhead{MA93\tablenotemark{c}} & \colhead{Martayan07\tablenotemark{d}} & \colhead{Martayan10\tablenotemark{e}} 
& \colhead{Evans04\tablenotemark{f}} & \colhead{Evans06\tablenotemark{g}}
}
\startdata
O-star Total & 778 & \nodata & \nodata & \nodata & 107 & 13 \\
O-stars in OKP & 778 & \nodata & \nodata  & \nodata & 94 & 9 \\
Oe Total & 171 & \nodata & 1  & 11 & 2 & 2 \\
Oe in OKP & 171 & \nodata &  1 & 6 & 2 & 2 \\
OB Total & 1220 & \nodata &  \nodata & \nodata & 459 & 54 \\
OB in OKP & 1220 & \nodata & \nodata & \nodata & 239 & 30 \\
OBe Total & 447 & 461 & 14 & 23 & 29 & 16 \\
OBe in OKP & 447 & 250 & 3 & 10 & 21 & 11 \\
\hline
O-star Recovery & \nodata & \nodata  & \nodata & \nodata & 0.88 & 0.69 \\
Oe Recovery & \nodata & \nodata & 1.0 & 0.55 & 1.0 & 1.0 \\
OB Recovery & \nodata & \nodata & \nodata & \nodata & 0.52 & 0.56 \\
OBe Recovery & \nodata & 0.54 & 0.21 & 0.44 & 0.72 & 0.69 
\enddata
\tablenotetext{a}{
Recovery fractions of OKP stars in the given surveys are shown, based on the spectral types in the given survey, for OB stars of spectral type $\leq$ B2 and luminosity classes III -- V.  Stars of uncertain types are included if the range is consistent with these criteria. 
Here, OBe stars are included in the OB survey numbers.}
\tablenotetext{b}{\citet{OKP} survey, for $B\leq 15.21$, retaining 480 stars with no published spectral types as OB candidates.}
\tablenotetext{c}{\citet{Meyssonnier1993} slitless SMC OBe survey, for $V\leq 15.5$, selecting their parameters "spec"="em" or "em:" and excluding objects with an "Obj" classification or with H$\alpha$ classified as sharp, "Sp=c1".  MA93 do not provide spectral types.}
\tablenotetext{d}{\citet{Martayan2007} OBe spectroscopic survey of rich cluster NGC 330, for $V\leq 15.5$.}
\tablenotetext{e}{\citet{Martayan2010} OBe slitless survey of 84 SMC open clusters, for $B\leq 15.21$.}
\tablenotetext{f}{\citet{Evans2004} SMC spectroscopic survey, for \citet{Massey2002} $B_M\leq 15.21$ and $\leq15.21$ if no $B_M$ available.}
\tablenotetext{g}{\citet{Evans2006} spectroscopic survey of rich clusters NGC 330 and NGC 346, for $V\leq 15.5$.}
\label{tab:Completeness}
\end{deluxetable*}

\begin{deluxetable*}{lcl}
\tablecaption{Membership for Maximum CDF Sub-populations} 
\tablewidth{0pt}
\tablehead{
\colhead{Subsample} & \colhead{Total} & \colhead{Membership\tablenotemark{a}}
}
\startdata
Early O & 464 & Table~\ref{tab:Populations} rows 1, 4, 8 \\
\multicolumn{3}{l}{Subtracted from row 4: 6908, 19481, 41095, 44634, and  83235} \\
\multicolumn{3}{l}{Subtracted from row 8:  1600, 18871, 23352, 15380, 27712, 31574, 34457, 49580} \\
\hline
Late O & 552 & Table~\ref{tab:Populations} rows 2, 4, 8 \\
\multicolumn{3}{l}{Subtracted from row 4: 19481} \\
\multicolumn{3}{l}{Subtracted from row 8: 15380, 18871, 23352, 27712, and 31574} \\
\hline
B stars & 600 & Table~\ref{tab:Populations} rows 3, 4, 8 \\
\multicolumn{3}{l}{Subtracted from row 4: 7382, 10505, 11238, 16734, 22178, 38695, 38703, 41095, 53324, 55808, 56834, 60439, and 66160}\\
\multicolumn{3}{l}{Subtracted from row 8: 1600, 34457, and 49580}\\
\hline
Oe stars & 159 & Table~\ref{tab:Populations} rows 5, 7, 8 \\
\multicolumn{3}{l}{Subtracted from row 8: 1600, 8178, 11019, 11209, 12308, 12403, 13168, 13986, 14592, 15321, 15380, 18871, 21534, 23352, } \\
\multicolumn{3}{l}{27610, 27712, 28076, 31574, 34457, 49580, 54723, 56139, 58751, 60684, 64736, 68629} \\
\hline
Be stars & 370 & Table~\ref{tab:Populations} rows 6, 7, 8 
\enddata
\tablenotetext{a}{Each subsample consists of stars from the given rows of Table~\ref{tab:Populations}, omitting the listed stars.  These are removed since they do not meet the subsample's spectral type criteria or are binary stars that are already counted in another included row for the subsample.}
\label{tab:accounting}
\end{deluxetable*}

\clearpage
\bibliographystyle{aasjournal}{}
\bibliography{Dallas}

\end{document}